\documentstyle[twoside,fleqn,espcrc2,epsfig]{article}
\newcommand{\third}{\mbox{\small $\frac{1}{3}$}}
\newcommand{\ninth}{\mbox{\small $\frac{1}{9}$}}
\newcommand{\cO}{{\cal O}}
\newcommand{\Dd}[1]{\mbox{
  \parbox[b]{0cm}{$D$}\raisebox{1.7ex}{$\leftrightarrow$}$_{\!#1}$}}

\newcommand{\AmS}{{\protect\the\textfont2
  A\kern-.1667em\lower.5ex\hbox{M}\kern-.125emS}}

\newcommand{\sze}{\scriptsize}
 
\begin{document}
 
\title{\vspace{-3.65cm}
       {\normalsize DESY 98--116}    \\[-0.2cm]
       {\normalsize TPR--98--23}     \\[-0.2cm]
       {\normalsize HUB--EP--98/49}  \\[-0.2cm]
       {\normalsize September 1998}   \\
       \vspace{1.95cm}
Composite operators in lattice QCD: nonperturbative
renormalization\thanks{Talk given by M. G\"ockeler at Lat98,
                       Boulder, U.S.A.}}


\author{M. G\"ockeler\address{Institut f\"ur Theoretische Physik,
     Universit\"at Regensburg, D-93040 Regensburg, Germany},
R. Horsley\address{Institut f\"ur Physik, Humboldt-Universit\"at zu Berlin, 
       D-10115 Berlin, Germany}, 
H. Oelrich\address{Deutsches Elektronen-Synchrotron DESY and NIC,
       D-15735 Zeuthen, Germany},
H. Perlt\address{Institut f\"ur Theoretische Physik, Universit\"at 
      Leipzig, D-04109 Leipzig, Germany},
D. Petters$^{\rm c,}$\address{Institut f\"ur Theoretische Physik, 
      Freie Universit\"at Berlin, D-14195 Berlin, Germany},
P.E.L. Rakow$^{\rm a}$,
A. Sch\"afer$^{\rm a}$,
G.~Schierholz$^{\rm c,}$\address{Deutsches Elektronen-Synchrotron DESY,
      D-22603 Hamburg, Germany} and
A. Schiller$^{\rm d}$}
 
\begin{abstract}
We investigate the nonperturbative renormalization of composite
operators in lattice QCD restricting ourselves to operators
that are bilinear in the quark fields.
These include operators which are relevant to the calculation of 
moments of hadronic structure functions. The computations are based on
Monte Carlo simulations using quenched Wilson fermions.
\end{abstract}
 
\maketitle
 
\section{INTRODUCTION}

Lattice investigations of hadronic structure require the calculation of
hadronic matrix elements of composite operators. In general,
one has to convert the bare lattice operators into renormalized 
continuum operators by multiplication with the appropriate 
renormalization coefficient $Z$. The accuracy of perturbative 
calculations of these $Z$ factors remains uncertain, even if 
tadpole improvement \cite{lepmac} is used, as more than a 
one-loop computation is rarely available. Therefore a nonperturbative
calculation by Monte Carlo simulations appears to be an attractive
alternative \cite{marti}. 

We have performed such a computation for
operators that determine moments of hadronic structure functions.
Computing the $Z$ factors for a 
rather large range of renormalization scales we try to
find out at which scales (if at all) perturbative behavior 
sets in such that the multiplication with perturbative Wilson
coefficients makes sense. For further details and results see
Ref.~\cite{zpaper}. 

We use standard Wilson fermions in the quenched 
approximation. At $\beta = 6.0$ ($\beta = 6.2$) we work
on a $16^3 \times 32$ ($24^3 \times 48$) lattice.
For both $\beta$'s we have studied three values of the 
hopping parameter $\kappa$.

In this talk, we shall concentrate on one particular
example, namely the operator
\begin{equation} \label{opdef}
 \cO =
    \bar{q} \left( \gamma_4 \Dd{4} - \frac{1}{3} \sum_{i=1}^3
   \gamma_i \Dd{i} \right)  q \,, 
\end{equation}
($\cO_{v_{2,b}}$ in the notation of Ref.~\cite{zpaper})
whose hadronic matrix elements are proportional to the momentum
fraction $\langle x \rangle$ carried by the quarks. 
Working with Wilson fermions it is straightforward to write
down a lattice version of the above operator. One simply replaces the
continuum covariant derivative by its lattice analogue. 

\section{THE METHOD}

We follow the procedure proposed by Martinelli et al.~\cite{marti}.
It applies the definitions used in (continuum) perturbation theory
to vertex functions calculated by Monte Carlo simulations on the 
lattice. To be more specific, we calculate the quark-quark Green 
function (a matrix in color and Dirac space) with one insertion 
of the operator $\cO$ at momentum 
zero in the Landau gauge. Amputating the external quark legs
we obtain the vertex function $\Gamma(p)$ as a function of the quark
momentum $p$. Defining the renormalized vertex function by 
$ \Gamma_{\mathrm R} (p) = Z_q^{-1} Z \Gamma (p) $
we fix the renormalization constant $ Z $ by imposing the
condition
\begin{equation} \label{defz}
 \mbox{\small $\frac{1}{12}$} {\rm tr} \left( \Gamma_{\mbox{\sze R}} (p)
   \Gamma_{\mbox{\sze Born}} (p) ^{-1} \right) = 1 
\end{equation}
at $p^2 = \mu^2$, where $\mu$ is the renormalization scale. 
So we calculate $ Z $ from
\begin{equation} \label{calcz} 
 Z_q^{-1} Z \mbox{\small $\frac{1}{12}$} {\rm tr} \left( \Gamma (p)
   \Gamma_{\mbox{\sze Born}} (p) ^{-1} \right) = 1 
\end{equation}
with $p^2 = \mu^2$. Here $\Gamma_{\mbox{\sze Born}} (p) $ is the 
Born term in the 
vertex function of $\cO$ computed on the lattice, and
$Z_q$ denotes the quark field renormalization constant. 
The latter is calculated from the quark propagator $S(p)$:
\begin{equation}
 Z_q (p) = \frac{ {\rm tr} \left( - {\rm i} \sum_\lambda \gamma_\lambda 
           \sin (a p_\lambda) a  S^{-1} (p) \right) }
           {12 \sum_\lambda \sin^2 (a p_\lambda) } \,, 
\end{equation}
again at $p^2 = \mu^2$.

\section{PERTURBATIVE INPUT}

Eq.~(\ref{defz}) defines a renormalization scheme
of the momentum subtraction type, which we call MOM scheme.
In order to convert our results into the more popular 
$\overline{\mbox{MS}}$ scheme we have to perform a
finite renormalization. 
The corresponding renormalization constant 
$Z^{\overline{\mathrm MS}}_{\mathrm MOM}$
is computed in continuum
perturbation theory using dimensional regularization.

Neglecting quark masses we obtain in the Landau gauge
\begin{equation}
  \begin{array}{l} \displaystyle
Z^{\overline{\mathrm MS}}_{\mathrm MOM} =
    1 + \frac{g^2}{16 \pi^2} C_F \Bigg[ - \frac{31}{9}
  \\ \displaystyle \quad
     {} - \frac{\left( p_4^2 - \third ( p_1^2 + p_2^2 + p_3^2 ) \right)^2 }
       { 3 p^2 \left( p_4^2 + \ninth ( p_1^2 + p_2^2 + p_3^2 ) \right) }
          \Bigg] + O(g^4) \,,
   \end{array}
\end{equation}
where $C_F=4/3$ for the gauge group SU(3).
Because of the noncovariance of the condition (\ref{defz}),
$Z^{\overline{\mathrm MS}}_{\mathrm MOM}$
depends on the direction of the momentum $p$.

For the scale dependence, the renormalization group predicts
(at fixed bare parameters)
\begin{equation} \label{rengroup}
  \begin{array}{l} \displaystyle
 R (\mu,\mu_0) := \frac{Z (\mu)}{Z (\mu_0)}
  \\ \displaystyle \quad
  {} =  \exp \left \{ - \int_{\bar{g}(\mu_0^2)}^{\bar{g}(\mu^2)} 
    \mathrm{d}g \frac{\gamma(g)}{\beta(g)} \right \} 
\end{array}
\end{equation}
in terms of the running coupling $\bar{g}(\mu^2)$, the  $\beta$-function 
$\beta (g)$, and the anomalous dimension $\gamma (g)$. An analogous formula
describes the $\beta$ dependence at fixed renormalized quantities.

\section{RESULTS}

Calculating the necessary two- and three-point functions with the
help of momentum sources \cite{zpaper,harald,thesis} we achieve
small statistical errors with a moderate number of configurations.
After extrapolating to the chiral limit (linearly in $1/\kappa$)
we multiply by $ Z^{\overline{\mathrm MS}}_{\mathrm MOM} $.

For the presentation of our results we convert lattice units 
to physical units using $a^{-2} = 3.8$ GeV$^2$ ($\beta = 6.0$) 
and $a^{-2} = 7.0$ GeV$^2$ ($\beta = 6.2$)
as determined from the string tension \cite{scaling}. 
The $\Lambda$ parameter 
is taken to be $\Lambda_{\overline{\mathrm{MS}}} = 230$ MeV. 

The scale dependence of our $Z$'s should be described by the
renormalization group factor $R$ (see (\ref{rengroup})). 
Hence we divide our numerical results by this expression
(evaluated in two-loop approximation with $\mu_0^2 = 4$ GeV$^2$) 
and define $Z_{\mathrm{RGI}} = Z/R$. For $Z_{\mathrm{RGI}}$ 
we hope to obtain a $\mu$ independent answer, at least in a reasonable
window of $\mu$ values. There $\mu^2$ would be large
enough to allow for perturbative scaling behavior.
On the other hand, $\mu^2$ should be small enough to
avoid strong cut-off effects.
\begin{figure}
\vspace*{-0.8cm}
\epsfig{file=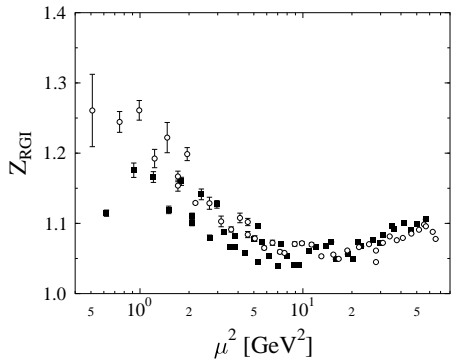,width=7.5cm} 
\vspace*{-1.5cm}
\caption{$Z_{\mathrm{RGI}}$ for the operator $\cO$ (Eq.~(\ref{opdef})).
         The open circles (filled squares) 
         represent the data for $\beta = 6.2$ ($\beta = 6.0$).
         The $\beta = 6.2$ data have been rescaled perturbatively
         to $\beta = 6.0$.}
\label{fig.r.v2b}  
\end{figure}

In Fig.~\ref{fig.r.v2b} we plot $Z_{\mathrm{RGI}}$
versus the renormalization scale $\mu^2$ for our operator $\cO$.
The errors are purely statistical.
A ``flat'' region seems to start at $\mu^2 \approx 5$ GeV$^2$.
Note that the $\beta = 6.0$ data and the 
(perturbatively rescaled) $\beta = 6.2$ data agree except for 
the lowest values of $\mu^2$. This indicates
that the observed $\mu$ dependence is physical.
For other operators, the results look qualitatively similar, although
in most cases a window starts only at $\mu^2 \approx 10$ GeV$^2$, 
where it is hard to believe that cut-off effects are negligible. 

\begin{figure}
\vspace*{-2.0cm}
\epsfig{file=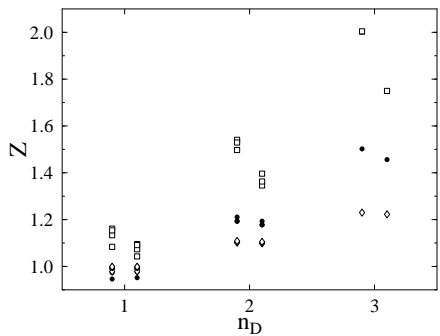,width=7.5cm} 
\vspace*{-2.0cm}
\caption{Collection of $Z$'s in the $\overline{\mathrm{MS}}$ 
         scheme versus the number of derivatives in the operators.
         The squares represent the nonperturbative results, the
         diamonds (filled circles) correspond to the values obtained 
         in lattice perturbation theory (with tadpole improvement).
         The $\beta = 6.0$ ($\beta = 6.2$) data are displaced to 
         the left (right).}  
\label{fig.2}  
\end{figure}

In Fig.~\ref{fig.2} we compare the nonperturbative values for
$Z$ in the $\overline{\mathrm{MS}}$ scheme with
(tadpole improved) one-loop lattice perturbation theory. We plot
the results for all operators with derivatives studied in 
Ref.~\cite{zpaper} at
$\mu^2 = a^{-2}$ versus $n_D$, the number of derivatives.
One-loop perturbation theory underestimates the 
increase of $Z$ with $n_D$. Tadpole improvement works in the
right direction for $n_D > 1$ without being quantitatively
satisfactory. However, the differences between the three methods
decrease as $\beta$ (and $\mu^2$) becomes larger.  

\section{SUMMARY}

In summary, we have performed a comprehensive study of 
nonperturbative renormalization
for various operators in the framework of lattice QCD.
Twist-2 operators appearing in unpolarized
(polarized) deep-inelastic scattering 
have been studied for all spins $\leq 4$ ($\leq 3$)~\cite{zpaper}.
The scale dependence of $Z$ for the operator
considered in this talk agrees with perturbative expectations
to a good approximation already for moderate values of $\mu^2$.
However, most of the other operators containing covariant
derivatives seem to approach perturbative scaling only for rather
large values of the renormalization scale where cut-off effects
might contaminate the results. 
The behavior at smaller $\mu^2$ could be nonperturbative physics, 
but at present we cannot exclude the possibility that it is faked, 
e.g., by finite-size effects. 
In any case, it might not always be sensible to
combine these nonperturbative
$Z$'s with perturbative Wilson coefficients using a scale of,
say, $\mu^2 = 4$ GeV$^2$. We have therefore started 
a nonperturbative calculation of the 
Wilson coefficients. First results are reported in Ref.~\cite{daniel}. 

\section*{ACKNOWLEDGEMENTS}
This work is supported by the Deut\-sche For\-schungs\-ge\-mein\-schaft and 
by BMBF. The numerical calculations were performed on the Quadrics 
computers at DESY-Zeuthen. We wish to thank the operating staff 
for their support.


\end{document}